\documentclass[aps,prd,amsmath,twocolumn,amssymbaps,showpacs]{revtex4-1}
\usepackage{graphicx}  
\usepackage{dcolumn}   
\usepackage{bm}        
\usepackage{amssymb}   
\usepackage{amsmath}   
\usepackage{lineno}
\usepackage{draftcopy}
\usepackage{relsize} 
\usepackage{xfrac}    
\usepackage{slashed}

\usepackage{color}
\definecolor{dgreen}{cmyk}{1.,0.,1.,0.2}        
\definecolor{orange}{cmyk}{0.,0.353,1.,0.}    

\usepackage[bookmarks]{hyperref}

\newcommand{\di}{{\rm d}}

\newcommand{\be}{\begin{equation}}
\newcommand{\ee}{\end{equation}}                                                                               
\newcommand{\bea}{\begin{eqnarray}}
\newcommand{\eea}{\end{eqnarray}}

\begin{document}
\title{Quarkyonic matter state of neutron stars}

\author{Gaoqing Cao}\email{caogaoqing@mail.sysu.edu.cn }
\affiliation{School of Physics and Astronomy, Sun Yat-sen University, Zhuhai 519088, China.} 

\date{\today}
\begin{abstract} 
This work extends our previous study of isospin symmetric quarkyonic matter to quarkyonic neutron matter which might be relevant to the inner cores of neutron stars. The vector-isovector $\rho$ mesons are introduced to the model mainly to account for isospin density interactions, just like $\omega$ meson for baryon density interactions. The modified Lagrangian still preserves approximate chiral symmetry which could be significantly broken at lower density. And new free parameters are fixed by adopting the experimental constraints on the symmetry energy and its slope at saturation density. Eventually, the pressure, mass-radius relation, and tidal deformability are explored in advance for the quarkyonic neutron stars. While the pressure and tidal deformability are well consistent with experimental and observational restrictions, the mass-radius relation is unable to reproduce the observed two solar mass of PSR J0740+6620.
\end{abstract}

\pacs{11.30.Qc, 05.30.Fk, 11.30.Hv, 12.20.Ds}

\maketitle

\section{Introduction}\label{introduction}
Recently, neutron stars are widely and heatedly discussed due to several improvements of astronomy detections and intriguing theoretical proposals. First of all, the terrestrial gravitational wave (GW) observatories unlocked a new gate to explore the properties of neutron stars through the binary mergers~\cite{LIGOScientific:2018cki,Annala:2017llu,De:2018uhw,Tews:2018iwm,Margalit:2017dij}. The emission of GW mainly depends on the time evolution of quadrupole moment, thus tidal deformability of neutron stars, closely related to the equation of state (EOS), could be extracted from the GW spectrum~\cite{LIGOScientific:2018cki,Annala:2017llu,De:2018uhw,Tews:2018iwm}. And the radii and maximum masses of neutron stars could be further constrained by combining GW with electromagnetic signals~\cite{Margalit:2017dij}. Second, the advanced detector NICER, installed aboard the International Space Station in 2017, could help to precisely measure the radii of neutron stars~\cite{Riley2021,Miller2021,Raaijmakers2021} and accordingly further constrain the hardness of the EOS. Finally, besides the color superconductivity phase inside the inner cores of neutron stars~\cite{Alford:2007xm}, the possibility of quarkyonic matter state has inspired new interests in an even wider community~\cite{Fukushima:2015bda,McLerran:2018hbz,Xia:2018cpy,Cao:2020byn} .
 
Quarkyonic matter is a state where baryons coexist with deconfined quarks at high baryon density~\cite{McLerran:2007qj}: low-momentum quarks are free or quasi-free but high-momentum ones are mainly confined inside baryons. In our previous work~\cite{Cao:2020byn}, we tried to develop a complete field model for quarkyonic matter by considering baryons, quarks, and mesons in the same level and consistently taking chiral symmetry breaking and restoration into account. For the isospin symmetric case, we fixed free parameters by fitting to the saturation properties of stable nuclei and thus obtained pressure-density relation is consistent with the experimental constraints~\cite{Danielewicz2002}. To complete the discussions, the study must be extended to isospin asymmetric case with neutron stars the most important natural correspondences. That would help us to check if the model is good enough for the whole isospin range on one hand or give hints on the missing physics on the other hand.

The short paper is arranged as follows. In Sec.\ref{Formulism}, we present the whole theoretical framework for quarkyonic neutron matter with the fundamental Lagrangian density and  state function given in Sec.\ref{Lagrangian} and the corresponding gap equations and thermodynamic quantities derived in Sec.\ref{gap}. In Sec.\ref{results}, we fix the new free parameters, carry out numerical calculations, and demonstrate the whole results. Finally, we briefly conclude in Sec.\ref{conclusion}.
\section{An effective model for quarkyonic neutron matter}\label{Formulism}

\subsection{Lagrangian and state function}\label{Lagrangian}
Previously, we constructed a field theoretical model for the isospin symmetric quark-baryonic (or quarkyonic) matter (QBM)~\cite{Cao:2020byn} by combining  the quark-meson model~\cite{Schaefer:2006ds} with the well-known Walecka model~\cite{Walecka}. By definition, quarkyonic matter belongs to the canonical ensemble where total densities are fixed but total energy could fluctuate~\cite{Duarte:2021tsx}. So the relevant chemical potentials should be determined according to the total densities and minimization of the Helmholtz free energy $F$ with respect to quark ratios in principle. For isospin symmetric QBM, it is natural to set the chemical potential relations $\mu_{\rm p}=\mu_{\rm n}, \mu_{\rm p}'=\mu_{\rm n}',$ and $\mu_{\rm u}=\mu_{\rm d}$. Here, $\mu_{\rm p/n}'$ reflects the blocking effect from quarks and thus should depend on $\mu_{\rm u/d}$~\cite{Cao:2020byn}. Then, the total baryon density and minimization of $F$ are enough to fix all the chemical potentials. For isospin asymmetric QBM, such as those in neutron stars, there are four independent chemical potentials in principle, $\mu_{\rm p}, \mu_{\rm n}, \mu_{\rm u}$, and $\mu_{\rm d}$. Similarly, they should be fixed by the constraints of total baryon and isospin densities and the corresponding minimization conditions. For neutron stars, the fraction of protons  was usually found to be small when beta equilibrium was taken into account~\cite{Li:2008gp}. So to simplify our discussions, we neglect the protons degrees of freedom but require the chemical equilibria,
\bea\label{ceq}
\mu_{\rm n}=2\mu_{\rm d}+\mu_{\rm u},\ \ \mu_{\rm d}=\mu_{\rm u}+\mu_{\rm e}
\eea
by following our previous scheme~\cite{Cao:2020byn}. One should note here that we choose to work in the grand canonical ensemble for two reasons: it is more convenient to depict the blocking effect of  quarks to nucleons and it better follows the philosophy of chiral symmetry breaking and restoration. As long as there is no real phase transition, the two ensembles should be equivalent.

Getting rid of the proton parts, the Lagrangian density for the quarkyonic neutron matter (QNM) can be modified from the previous one~\cite{Cao:2020byn} as
\begin{widetext}
\begin{eqnarray}\label{QH}
{\cal L}_{\rm q}\!\!&=&\!\!\bar{q}\Big[i\slashed{\partial}+\left({\mu_{\rm B}\over N_c}+{\mu_{\rm I}\over 2}\tau_3\right)\gamma^0-g_{q}\left(\sigma+i\gamma^5\boldsymbol{\tau\cdot\pi}\right)\Big]q,\nonumber\\
{\cal L}_{\rm n}\!\!&=&\!\!\bar{n}\Big[i\slashed{\partial}-\mu_{\rm n}\gamma^0-g_{\rm Ns}\left(\sigma-i\gamma^5{\pi^0}\right)-g_{\rm N\omega}\slashed{\omega}+g_{\rm N\rho}\left({\slashed{\rho}}^3-\gamma^5\slashed{A}^3\right)\Big]n,\nonumber\\
{\cal L}_{\rm M}\!\!&=&\!\!{1\over2}\left(\partial_\mu\sigma\partial^\mu\sigma+D_\mu\boldsymbol{\pi}\boldsymbol{\cdot}D^\mu\boldsymbol{\pi}\right)-{\lambda\over4}\left(\sigma^2+\boldsymbol{\pi\cdot\pi}-\upsilon^2\right)^2+c~\sigma-{1\over4}({\omega}_{\mu\nu}{\omega}^{\mu\nu}+\boldsymbol{\rho}_{\mu\nu}\boldsymbol{\cdot}\boldsymbol{\rho}^{\mu\nu}+{\bf{A}}_{\mu\nu}\boldsymbol{\cdot}  {\bf{A}}^{\mu\nu})\nonumber\\
&&+{1\over2}g_{\rm s\omega}\left(\sigma^2+\boldsymbol{\pi\cdot\pi}-h_\omega^2\right)\omega_\mu\omega^\mu+{1\over2}g_{\rm s\rho}\left(\sigma^2+\boldsymbol{\pi\cdot\pi}-h_\rho^2\right)\left(\boldsymbol{\rho}_\mu\boldsymbol{\cdot}\boldsymbol{\rho}^\mu+ {\bf{A}}_\mu\boldsymbol{\cdot}  {\bf{A}}^\mu\right),
\end{eqnarray}
\end{widetext}
where $\mu_{\rm n}\equiv\mu_{\rm B}-{\mu_{\rm I}\over 2}$ with $\mu_{\rm B}$ and ${\mu_{\rm I}}$ the baryon and isospin chemical potentials, respectively.
Here, the quantum fields are defined as the following: $q(x)=(u(x),d(x))^T$ denotes the two-flavor quark field with color degrees of freedom $N_c=3$, $n(x)$ is the neutron field {\it outside} the Fermi spheres of quarks if exist; $\sigma(x)$ and $\boldsymbol{\pi}(x)$ are the scalar and pseudoscalar mesons, respectively; $\omega_\mu$ is a vector-isoscalar meson; and $\boldsymbol{\rho}_\mu$ and ${\bf{A}}_\mu$ are vector-isovector and axial-isovector mesons, respectively. Compared to our naive presentation of the Lagrangian in the previous work, we note that the interactions involving $\omega_\mu$ are not necessarily degenerate with those involving $\boldsymbol{\rho}_\mu$ since they belong to different isospin groups, though their masses are quite close to each other. The isospin matrices are   
$$\tau=\left(1,{\tau_x-i\tau_y\over\sqrt{2}},{\tau_x+i\tau_y\over\sqrt{2}},\tau_z\right)$$
with $\tau_{x},\tau_{y},$ and $\tau_{z}$ the Pauli matrices in flavor space. The  derivative operators are defined as $D_0=\partial_0\mp i{\mu_{\rm I}}$ for the charged $\pi^\pm,\boldsymbol{\rho}^\pm_\mu,$ and ${\bf A}_\mu^\pm$, and $D_\mu=\partial_\mu$ for the others. 

In the vacuum, only $\langle\sigma\rangle$ is expected to be nonzero and the  thermodynamic potential is given by~\cite{Schaefer:2006ds}
 \begin{eqnarray}
\Omega_v={\lambda\over4}\left(\langle\sigma\rangle^2+{\langle\boldsymbol{\pi}\rangle\cdot\langle\boldsymbol{\pi}\rangle}-\upsilon^2\right)^2-c~\langle\sigma\rangle
\end{eqnarray}
in mean field approximation. We now turn to compute thermodynamics at finite temperature and chemical potentials, where both quarks and neuntrons will give contributions. For the study of neutron stars, we confine ourselves to the case $\mu_{\rm B}>-\mu_{\rm I}/2>0$, which might involve two kinds of vector condensations, $\langle\omega_0\rangle$ and $\langle\rho_0^3\rangle$. Then, the vacuum term would be modified to
 \begin{eqnarray}
\Omega_v&=&{\lambda\over4}\left(\langle\sigma\rangle^2+{\langle\boldsymbol{\pi}\rangle\cdot\langle\boldsymbol{\pi}\rangle}-\upsilon^2\right)^2-c~\langle\sigma\rangle\nonumber\\
&&-{g_{\rm s\omega}\over2}(\langle\sigma\rangle^2-h_\omega^2)\langle\omega_0\rangle^2-{g_{\rm s\rho}\over2}(\langle\sigma\rangle^2-h_\rho^2)\langle\rho_0^3\rangle^2.
\end{eqnarray}
It is easy to work out the thermodynamic potentials for the quark and neutron parts according to the imaginary-time field theory~\cite{Kapusta2006}, we have
\begin{widetext}
\begin{eqnarray}
\Omega_{\rm q}&=&-2N_cT\!\!\sum_{l,t=\pm}\!\int{d^3p\over(2\pi)^3}\ln\Big(1+e^{-[E_{\rm q}({\bf p})+l\left({\mu_{\rm B}\over N_c}+t{\mu_{\rm I}\over 2}\right)]/T}\Big),\\
\Omega_{\rm n}&=&-{1\over2}g_{\rm s\omega}(\langle\sigma\rangle^2-h_\omega^2)\langle\omega_0\rangle^2-{1\over2}g_{\rm s\rho}(\langle\sigma\rangle^2-h_\rho^2)\langle\rho_0^3\rangle^2-2T\sum_{l=\pm}\int{d^3p\over(2\pi)^3}\ln\Big(1+e^{-[E_{\rm n}({\bf p})+l(\mu_{\rm n}-(g_{\rm N\omega} \langle\omega_0\rangle-g_{\rm N\rho}\langle\rho_0^3\rangle))]/T}\Big)\nonumber\\
&&+2T\sum_{l=\pm}\int{d^3p\over(2\pi)^3}\ln\Big(1+e^{-[E_{\rm n}({\bf p})+l(\mu_{\rm n}'-(g_{\rm N\omega} \langle\omega_0\rangle-g_{\rm N\rho}\langle\rho_0^3\rangle))]/T}\Big).\label{Omegan}
\end{eqnarray}
\end{widetext}
Here, $\mu_{\rm n}'$ is the chemical potential for the blocked neutron sea, and the dispersion relations are 
 \begin{eqnarray}
E_{\rm q/n}({\bf p})=\sqrt{{\bf p}^2+m_{\rm q/n}^2}
 \end{eqnarray}
with $m_{\rm q}=g_{\rm q}\langle\sigma\rangle$ and $m_{\rm n}=g_{\rm Ns}\langle\sigma\rangle$, respectively.
The vector condensates are subject to the physical constraint $0\leq (g_{\rm N\omega} \langle\omega_0\rangle-g_{\rm N\rho}\langle\rho_0^3\rangle)\leq\mu_{\rm n}$, which means that the neutron chemical potential is reduced by $\langle\omega_0\rangle$ and $\langle\rho_0^3\rangle$ but never changes sign.

The  crucial step here is to implement the quarkyonic picture in the momentum space, in which the interior of the Fermi sea is filled up by quarks while the nucleons reside in an outside shell~\cite{McLerran:2007qj,McLerran:2008ua}. As one can tell in Eq.\eqref{Omegan}, $\Omega_{\rm n}$ is obtained by subtracting the supposed inner contribution (with ${\mu_{\rm n}'}$) out of the naive total one (with $\mu_{\rm n}$), that is, the neutrons only exist between the Fermi spheres stretched by ${\mu_{\rm n}'}$ and $\mu_{\rm n}$. To carry out further calculations, we should find an appropriate scheme to determine the form of ${\mu_{\rm n}'}$. Based on comparison of the kinetic energy of a neutron with that of $2u+d$ quarks, we choose the $\mu_{\rm n}$-linear form:
\begin{eqnarray}\label{muPB1}
\mu_{\rm n}'=\mu_{\rm n}-(N_c m_{\rm q}-m_{\rm n}).
\end{eqnarray} 
Compared to the non-linear form, the advantage of this scheme is that the blocked neutron density is in the same form as the naive neutron density~\cite{Cao:2020byn}, except for different chemical potentials, according to the thermodynamic relation $n_{\rm n}=-\partial\Omega_{\rm n}/\partial\mu_{\rm n}$.

\subsection{Gap equations and thermodynamics}\label{gap}
In mean field approximation, the total thermodynamic potential is then $\Omega=\Omega_{\rm v}+\Omega_{\rm q}+\Omega_{\rm n}$ and the gap equations can be obtained from the extremal conditions $\partial\Omega/\partial X=0\ \ (X=\langle\omega_0\rangle,\langle\rho_0^3\rangle, \langle\sigma\rangle)$ as
\begin{widetext}
\begin{eqnarray}
&&\!\!\!\!\langle\omega_0\rangle=-{2}\sum_{l=\pm}\int{d^3p\over(2\pi)^3}{l{g_{\rm N\omega}\over g_{\rm s\omega}(\langle\sigma\rangle^2-h_\omega^2)}\over 1+e^{[E_{\rm n}({\bf p})+l(\mu_{\rm n}-(g_{\rm N\omega} \langle\omega_0\rangle-g_{\rm N\rho}\langle\rho_0^3\rangle))]/T}}+{2}\sum_{l=\pm}\int{d^3p\over(2\pi)^3}{l{g_{\rm N\omega}\over g_{\rm s\omega}(\langle\sigma\rangle^2-h_\omega^2)}\over 1+e^{[E_{\rm n}({\bf p})+l({\mu_{\rm n}'}-(g_{\rm N\omega}\langle\omega_0\rangle-g_{\rm N\rho}\langle\rho_0^3\rangle))]/T}},\label{omega0}\\
&&\!\!\!\!\langle\rho_0^3\rangle={2}\sum_{l=\pm}\int{d^3p\over(2\pi)^3}{l{g_{\rm N\rho}\over g_{\rm s\rho}(\langle\sigma\rangle^2-h_\rho^2)}\over 1+e^{[E_{\rm n}({\bf p})+l(\mu_{\rm n}-(g_{\rm N\omega} \langle\omega_0\rangle-g_{\rm N\rho}\langle\rho_0^3\rangle))]/T}}-{2}\sum_{l=\pm}\int{d^3p\over(2\pi)^3}{l{g_{\rm N\rho}\over g_{\rm s\rho}(\langle\sigma\rangle^2-h_\rho^2)}\over 1+e^{[E_{\rm n}({\bf p})+l({\mu_{\rm n}'}-(g_{\rm N\omega}\langle\omega_0\rangle-g_{\rm N\rho}\langle\rho_0^3\rangle))]/T}},\label{rho0}\\
&&\!\!\!\!{\lambda}\left(\langle\sigma\rangle^2-\upsilon^2\right)\langle\sigma\rangle-c-\langle\sigma\rangle(g_{\rm s\omega}\langle\omega_0\rangle^2+g_{\rm s\rho}\langle\rho_0^3\rangle^2)+2N_c\sum_{l,t=\pm}\int{d^3p\over(2\pi)^3}{{g_{q}m_{\rm q}/ E_{\rm q}({\bf p})}\over1+e^{[E_{\rm q}({\bf p})+l\left({\mu_{\rm B}\over N_c}+t{\mu_{\rm I}\over 2}\right)]/T}}\nonumber\\
&&\!\!\!\!+2\sum_{l=\pm}\int{d^3p\over(2\pi)^3}{{g_{\rm Ns}m_{\rm n}/E_{\rm n}({\bf p})}\over 1+e^{[E_{\rm n}({\bf p})+l(\mu_{\rm n}-(g_{\rm N\omega} \langle\omega_0\rangle-g_{\rm N\rho}\langle\rho_0^3\rangle))]/T}}-2\sum_{l=\pm}\int{d^3p\over(2\pi)^3}{{g_{\rm Ns}m_{\rm n}/E_{\rm n}({\bf p})}+l\,(g_{\rm Ns}-g_{\rm q}N_{\rm c})\over 1+e^{[E_{\rm n}({\bf p})+l({\mu_{\rm n}'}-(g_{\rm N\omega}\langle\omega_0\rangle-g_{\rm N\rho}\langle\rho_0^3\rangle))]/T}}=0.\label{sigma}
\end{eqnarray}

Furthermore, the baryon number, isospin number and entropy densities can be derived directly according to the thermodynamic relationships $n_{\rm B}=-\partial\Omega/\partial\mu_{\rm B}, n_{\rm I}=-\partial\Omega/\partial\mu_{\rm I}$ and $s=-\partial\Omega/\partial T$ as
\begin{eqnarray}
n_{\rm B}&=&-2\sum_{l,t=\pm}\int{d^3p\over(2\pi)^3}l{1\over 1+e^{[E_{\rm q}({\bf p})+l\left({\mu_{\rm B}\over N_c}+t{\mu_{\rm I}\over 2}\right)]/T}}-2\sum_{l=\pm}\int{d^3p\over(2\pi)^3}l\left({1\over 1+e^{[E_{\rm n}({\bf p})+l({\mu_{\rm n}}-(g_{\rm N\omega}\langle\omega_0\rangle-g_{\rm N\rho}\langle\rho_0^3\rangle))]/T}}\right.\nonumber\\
&&\left.\ \ \ \ \ -{1\over 1+e^{[E_{\rm n}({\bf p})+l({\mu_{\rm n}'}-(g_{\rm N\omega}\langle\omega_0\rangle-g_{\rm N\rho}\langle\rho_0^3\rangle))]/T}}\right),\label{Bnumber}\\
n_{\rm I}&=&-N_c\sum_{l,t=\pm}\int{d^3p\over(2\pi)^3}l\,t{1\over 1+e^{[E_{\rm q}({\bf p})+l\left({\mu_{\rm B}\over N_c}+t{\mu_{\rm I}\over 2}\right)]/T}}+\sum_{l=\pm}\int{d^3p\over(2\pi)^3}l\left({1\over 1+e^{[E_{\rm n}({\bf p})+l({\mu_{\rm n}}-(g_{\rm N\omega}\langle\omega_0\rangle-g_{\rm N\rho}\langle\rho_0^3\rangle))]/T}}\right.\nonumber\\
&&\left.\ \ \ \ \ -{1\over 1+e^{[E_{\rm n}({\bf p})+l({\mu_{\rm n}'}-(g_{\rm N\omega}\langle\omega_0\rangle-g_{\rm N\rho}\langle\rho_0^3\rangle))]/T}}\right),\label{Inumber}\\
s&=&2N_c\sum_{l,t=\pm}\int{d^3p\over(2\pi)^3}\left(\ln\Big(1+e^{-[E_{\rm q}({\bf p})+l\left({\mu_{\rm B}\over N_c}+t{\mu_{\rm I}\over 2}\right)]/T}\Big)+{E_{\rm q}({\bf p})+l\left({\mu_{\rm B}\over N_c}+t{\mu_{\rm I}\over 2}\right)\over T\left(1\!+\!e^{[E_{\rm q}({\bf p})+l\left({\mu_{\rm B}\over N_c}+t{\mu_{\rm I}\over 2}\right)]/T}\right)}\right)\nonumber\\
&&+2\sum_{l=\pm}\int{d^3p\over(2\pi)^3}\left(\ln\Big(1+e^{-[E_{\rm n}({\bf p})+l(\mu_{\rm n}-(g_{\rm N\omega} \langle\omega_0\rangle-g_{\rm N\rho}\langle\rho_0^3\rangle))]/T}\Big)+{E_{\rm n}({\bf p})\!+\!l(\mu_{\rm n}-(g_{\rm N\omega} \langle\omega_0\rangle-g_{\rm N\rho}\langle\rho_0^3\rangle))\over T\left(1\!+\!e^{[E_{\rm n}({\bf p})+l(\mu_{\rm n}-(g_{\rm N\omega} \langle\omega_0\rangle-g_{\rm N\rho}\langle\rho_0^3\rangle))]/T}\right)}\right.\nonumber\\
&&\left.\ \ \ \ \ -\ln\Big(1\!+\!e^{-[E_{\rm n}({\bf p})+l({\mu_{\rm n}'}-(g_{\rm N\omega} \langle\omega_0\rangle-g_{\rm N\rho}\langle\rho_0^3\rangle))]/T}\Big)-{E_{\rm n}({\bf p})\!+\!l({\mu_{\rm n}'}\!-\!(g_{\rm N\omega} \langle\omega_0\rangle-g_{\rm N\rho}\langle\rho_0^3\rangle))\over T\left(1\!+\!e^{[E_{\rm n}({\bf p})+l({\mu_{\rm n}'}-(g_{\rm N\omega} \langle\omega_0\rangle-g_{\rm N\rho}\langle\rho_0^3\rangle))]/T}\right)}\right).
\end{eqnarray}
Thus, the energy density of the QNM is found to be 
\begin{eqnarray}
\epsilon&\equiv&\Omega+\mu_{\rm B}n_{\rm B}+\mu_{\rm I}n_{\rm I}+sT-\epsilon_0\nonumber\\
&=&{\lambda\over4}\Big(\langle\sigma\rangle^2\!-\!\upsilon^2\Big)^2\!-\!c~\langle\sigma\rangle+{1\over2}g_{\rm s\omega}(\langle\sigma\rangle^2\!-\!h_\omega^2)\langle\omega_0\rangle^2+{1\over2}g_{\rm s\rho}(\langle\sigma\rangle^2\!-\!h_\rho^2)\langle\rho_0^3\rangle^2\!+\!2\sum_{l,t=\pm}\int{d^3p\over(2\pi)^3}{N_cE_{\rm q}({\bf p})\over1\!+\!e^{[E_{\rm q}({\bf p})+l\left({\mu_{\rm B}\over N_c}+t{\mu_{\rm I}\over 2}\right)]/T}}\nonumber\\
&&+2\sum_{l=\pm}\int{d^3p\over(2\pi)^3}\left({E_{\rm n}({\bf p})\over1\!+\!e^{[E_{\rm n}({\bf p})+l(\mu_{\rm n}-(g_{\rm N\omega} \langle\omega_0\rangle-g_{\rm N\rho}\langle\rho_0^3\rangle))]/T}}-{E_{\rm n}({\bf p})\!-\!l\,(N_c m_{\rm q}-m_{\rm n})\over 1+e^{[E_{\rm n}({\bf p})+l({\mu_{\rm n}'}-(g_{\rm N\omega} \langle\omega_0\rangle-g_{\rm N\rho}\langle\rho_0^3\rangle))]/T}}\right)-\epsilon_0,\label{energy}
\end{eqnarray}
where $\epsilon_0$ is the zero-point energy that must be excluded so that the detectable energy $\epsilon$ vanishes in the vacuum. Note that the gap equations Eqs.\eqref{omega0} and \eqref{rho0} has been used to get the final expression.

At zero temperature, the explicit forms of Eqs.\eqref{Bnumber} and \eqref{Inumber} for the baryon and isospin densities become
\begin{eqnarray}
n_{\rm B}&=&n_{\rm B}^{\rm q}+n_{\rm B}^{\rm n}\equiv {p_{\rm uF}^3+p_{\rm dF}^3\over3\pi^2}+{p_{\rm nF}^3-p_{\rm nF}'^3\over3\pi^2},\label{Bdensity}\\
n_{\rm I}&=&n_{\rm I}^{\rm q}+n_{\rm I}^{\rm n}\equiv N_c {p_{\rm uF}^3-p_{\rm dF}^3\over6\pi^2}-{p_{\rm nF}^3-p_{\rm nF}'^3\over6\pi^2},\label{Idensity}
\end{eqnarray}
where $p_{\rm u/d\,F}$ are the Fermi momenta of the occupied quark flavors, and $p_{\rm nF}$ and $p_{\rm nF}'$ are the Fermi momenta of the initially occupied and Pauli-blocked neutron states. The Fermi momenta are related to the chemical potentials through the Fermi energies as
\begin{eqnarray}
&&E_{\rm u/d\,F}\equiv E_{\rm q}(p_{\rm u/d\,F})={\mu_{\rm B}\over N_c}\pm{\mu_{\rm I}\over2},\ E_{\rm nF}\equiv E_{\rm n}(p_{\rm tF})=\mu_{\rm n}-(g_{\rm N\omega} \langle\omega_0\rangle-g_{\rm N\rho}\langle\rho_0^3\rangle),\nonumber\\ 
&&E_{\rm nF}'\equiv E_{\rm n}(p_{\rm tF}')={\mu_{\rm n}'}-(g_{\rm N\omega} \langle\omega_0\rangle-g_{\rm N\rho}\langle\rho_0^3\rangle).
\end{eqnarray}
As in the previous study~\cite{Cao:2020byn}, the momentum integrations involved in the gap equations Eqs.(\ref{omega0}-\ref{sigma}) and energy density Eq.(\ref{energy}) can be carried out explicitly with the help of Fermi momenta as
\begin{eqnarray}
0&=&g_{\rm s\omega}(\langle\sigma\rangle^2-h_\omega^2)\langle\omega_0\rangle-g_{\rm N\omega}n_{\rm B}^{\rm n},\label{omega01}\\
0&=&g_{\rm s\rho}(\langle\sigma\rangle^2-h_\rho^2)\langle\rho_0^3\rangle-2g_{\rm N\rho}n_{\rm I}^{\rm n},\label{rho01}\\
0&=&{\lambda}\left(\langle\sigma\rangle^2-\upsilon^2\right)\langle\sigma\rangle-c-\langle\sigma\rangle(g_{\rm s\omega}\langle\omega_0\rangle^2+g_{\rm s\rho}\langle\rho_0^3\rangle^2)+{g_{\rm Ns}m_{\rm n}\over{2\pi^2}}\Delta\left[E_{\rm nF}p_{\rm nF}-m_{\rm n}^2\ln\Big({E_{\rm nF}+p_{\rm nF}\over m_{\rm n}}\Big)\right]\nonumber\\
&&+N_c{g_{q}m_{\rm q}\over{2\pi^2}}\sum_{t=u,d}\left[E_{\rm tF}p_{\rm tF}-m_{\rm q}^2\ln\Big({E_{\rm tF}+p_{\rm tF}\over m_{\rm q}}\Big)\right]+(g_{\rm Ns}-g_{\rm q}N_{\rm c}){p_{\rm nF}'^3\over3\pi^2},\\
\epsilon&=&{\lambda\over4}\left(\langle\sigma\rangle^2-\upsilon^2\right)^2-c~\langle\sigma\rangle+{1\over2}g_{\rm s\omega}(\langle\sigma\rangle^2\!-\!h_\omega^2)\langle\omega_0\rangle^2+{1\over2}g_{\rm s\rho}(\langle\sigma\rangle^2\!-\!h_\rho^2)\langle\rho_0^3\rangle^2\nonumber\\
&&+{1\over8\pi^2}\Delta\left[2E_{\rm nF}^3p_{\rm nF}\!-\!m_{\rm n}^2E_{\rm nF}p_{\rm nF}\!-\!m_{\rm n}^4\ln\Big({E_{\rm nF}\!+\!p_{\rm nF}\over m_{\rm n}}\Big)\right]\nonumber\\
&&+{N_c\over8\pi^2}\sum_{t=u,d}\left[2E_{\rm tF}^3p_{\rm tF}-m_{\rm q}^2E_{\rm tF}p_{\rm tF}-m_{\rm q}^4\ln\Big({E_{\rm tF}+p_{\rm tF}\over m_{\rm q}}\Big)\right]-(N_c m_{\rm q}-m_{\rm n}){p_{\rm nF}'^3\over3\pi^2}-\epsilon_0,
\end{eqnarray}
\end{widetext}
where the symbol $"\Delta"$ means excluding the corresponding one with $n\rightarrow n'$ for the energy and momentum. 

\section{Numerical results}\label{results}
\begin{figure}[!htb]
	\centering
	\includegraphics[width=0.38\textwidth]{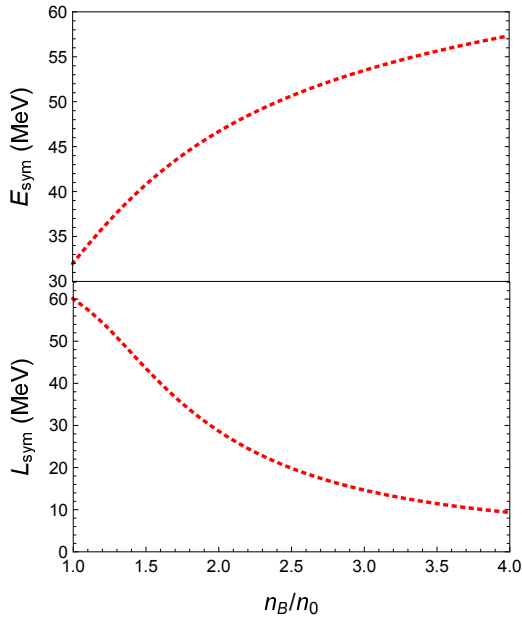}
	\caption{The symmetry energy $E_{\rm sym}(n_{\rm B})$ (upper panel) and its slope $L_{\rm sym}(n_{\rm B})\equiv3n_0{\partial E_{\rm sym}(n_{\rm B})\over\partial n_{\rm B}}$ (lower panel) as functions of the baryon density $n_{\rm B}$ for the quarkyonic neutron matter.}\label{ELsym}
\end{figure}

Most of the model parameters were fixed in our previous work already~\cite{Cao:2020byn}, that is, 
\bea
&&\lambda=30.56,\ \upsilon=89.59\,{\rm MeV},\ c=(121.0\,{\rm MeV})^3,\nonumber\\ 
&&g_{\rm N\omega}=7.232,\ g_{\rm s\omega}=81.06,\ h_\omega=293.5\,{\rm MeV}.
\eea
So here we only need to fix the new parameters relevant to $\rho$ mesons, that is, $g_{\rm N\rho}, g_{\rm s\rho}$, and $h_\rho$, the latter two of which are further connected by the vacuum mass $g_{\rm s\rho}(f_\pi^2-h_\rho^2)=(775\,{\rm MeV})^2$. Then, by recalling constraints given in the review Ref.~\cite{Oertel:2016bki}, $E_{\rm sym}(n_0)=31.6\pm2.7~{\rm MeV}$ and $L_{\rm sym}(n_0)=58.7\pm28.1~{\rm MeV}$, we will determine these parameters by requiring $E_{\rm sym}(n_0)=32~{\rm MeV}$ and $L_{\rm sym}(n_0)=60~{\rm MeV}$ for QNM at saturation density. In neutron stars, for a given $\mu_{\rm B}$, the variable $\mu_{\rm I}$ is actually fixed by chemical equilibria Eq.\eqref{ceq} and electric charge neutrality through
\begin{eqnarray}
2{p_{\rm uF}^3\over3\pi^2}-{p_{\rm dF}^3\over3\pi^2}+{\mu_{\rm I}^3\over3\pi^2}=0,
\end{eqnarray}
where the electric chemical potential of electrons is $-\mu_{\rm I}$ and we have neglected electron mass. Eventually, the fitting gives the new parameters as the following:
\bea
 g_{\rm N\rho}=3.491,\ g_{\rm s\rho}=92.49,\ h_\rho=446.5\,{\rm MeV},
\eea
and the associated evolutions of $E_{\rm sym}(n_{\rm B})$ and $L_{\rm sym}(n_{\rm B})$ are illustrated in Fig.\ref{ELsym}

\begin{figure}[!htb]
	\centering
	\includegraphics[width=0.42\textwidth]{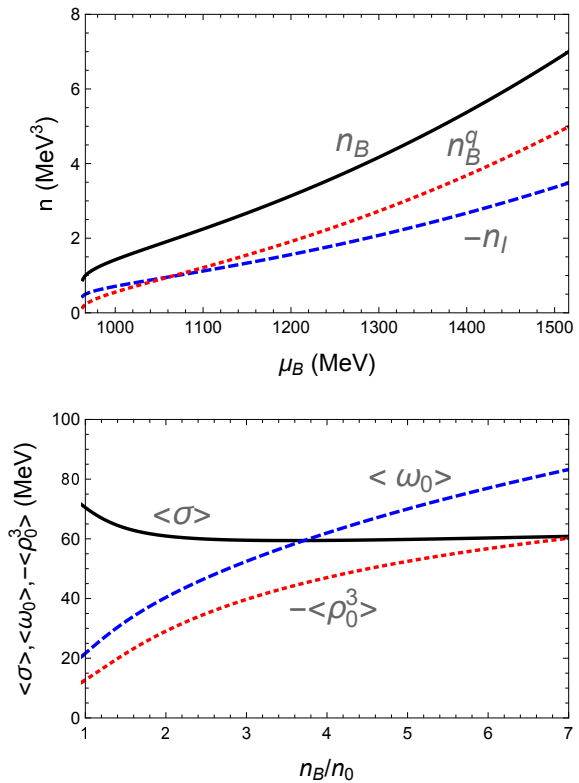}
	\caption{Upper panel: the densities $n_{\rm B}, n_{\rm I}$, and $n_{\rm B}^{\rm q}$ as functions of the baryon chemical potential $\mu_{\rm B}$; lower panel: the order parameters $\langle\sigma\rangle, \langle\omega_0\rangle$, and $\langle\rho_0^3\rangle$ as functions of the corresponding baryon density $n_{\rm B}$.}\label{dens_cond}
\end{figure}

\begin{figure}[!htb]
	\centering
	\includegraphics[width=0.42\textwidth]{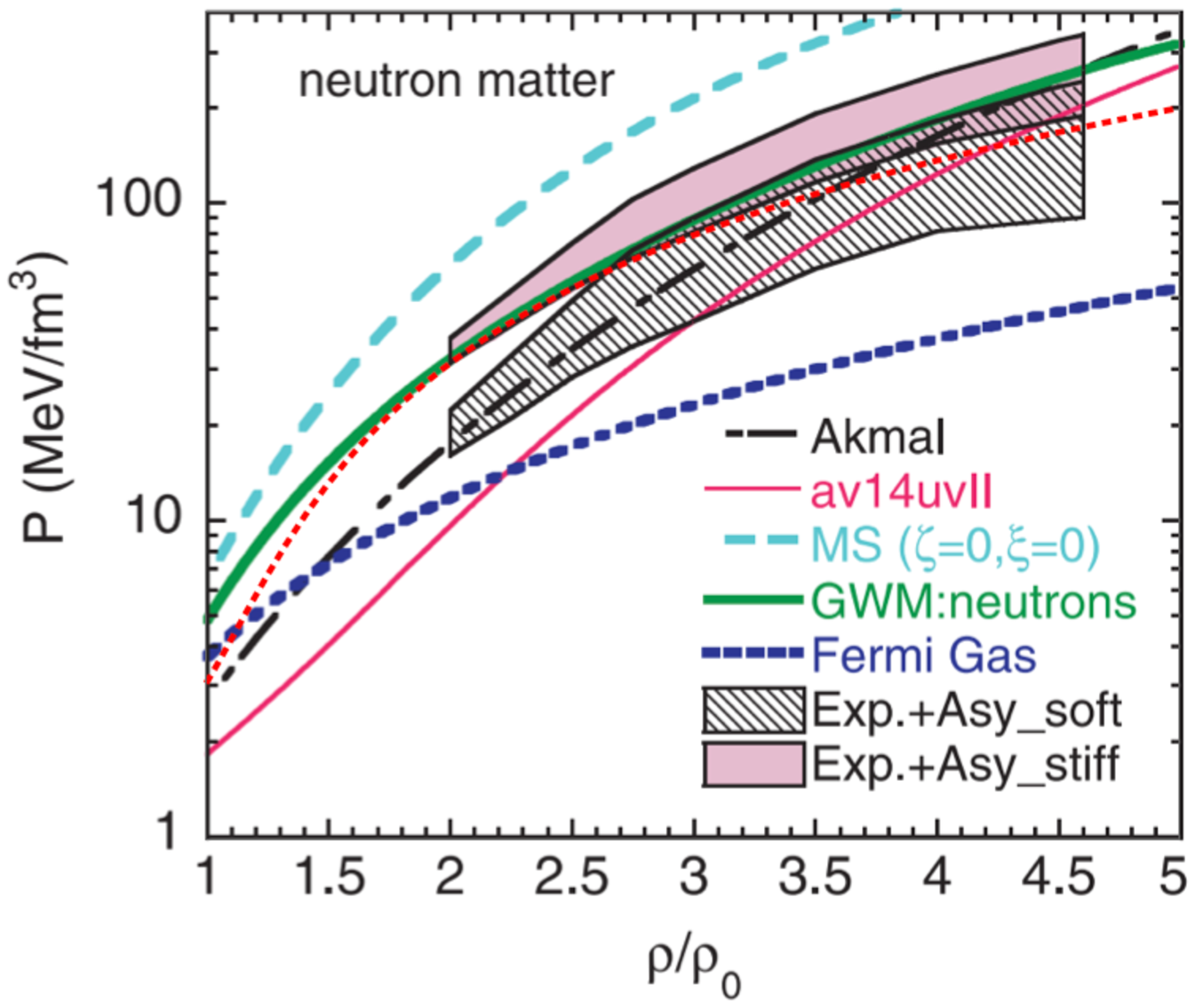}
	\includegraphics[width=0.38\textwidth]{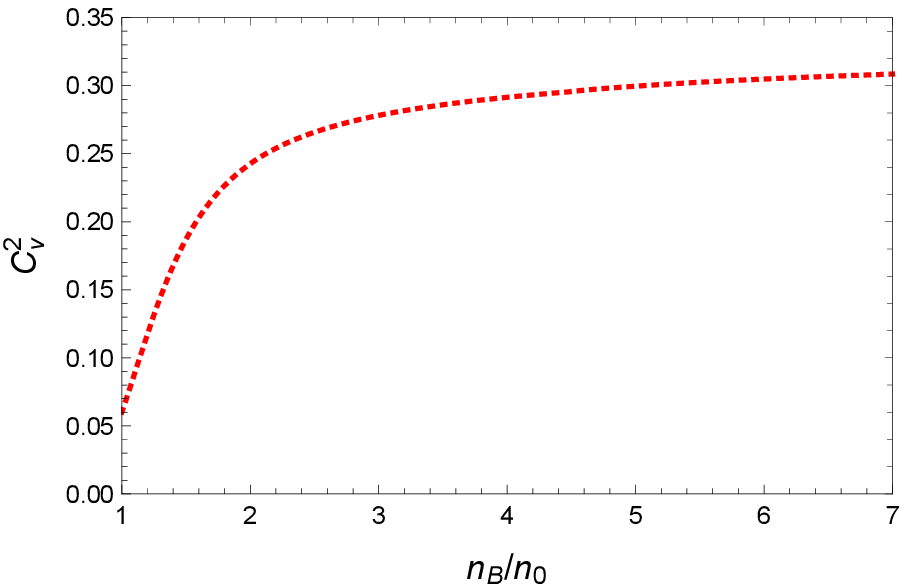}
	\caption{Pressure (red dotted in upper panel) and sound velocity (lower panel) as functions of baryon density $\rho$ or $n_{\rm B}$. The upper panel is adjusted from the plot given in Ref.~\cite{Danielewicz2002}.}\label{Pressure}
\end{figure}

Next, by increasing the baryon chemical potential $\mu_{\rm B}$, we explore the features of the densities and order parameters. As shown in the upper panel of Fig.\ref{dens_cond}, the baryon density and the ratio of quarks increase with $\mu_{\rm B}$, and we find $n_{\rm B}\approx-2n_{\rm I}$ as should be according to the definition of QNM. In the lower panel, correct responses of the order parameters to the densities are found: $\langle\sigma\rangle$ decreases with baryon densities due to chiral symmetry restoration, though only partially here; and $\langle\omega_0\rangle$ and $|\langle\rho_0^3\rangle|$ increase with neutron density.
Moreover, we show the pressure and square of sound velocity $C_{\rm v}^2=\partial P/\partial \epsilon$ as functions of the baryon density $n_{\rm B}$ in Fig.\ref{Pressure}. We are glad to see that the pressure is consistent with the experimental constraints very well~\cite{Danielewicz2002}, neither too soft nor too stiff. But the square of sound velocity gradually approaches the free quark limit $1/3$ from below without developing any peak structure, in contrast to other studies~\cite{McLerran:2018hbz,Kojo:2021wax} involving both nucleons and quarks (see the review Ref.~\cite{Baym:2017whm}). We note that there is no true phase transition in the considered region of baryon density, which thus justifies the application of the grand canonical ensemble picture to the density most relevant to neutron stars.

Finally, we apply the energy density and pressure obtained above to explore the observational properties of quarkyonic neutron stars, mainly the mass-radius relation.  For ordinary isotropic neutron stars, gravity is the reason why they can exist as bound objects and functions through the well-known Tolman–Oppenheimer–Volkov (TOV) equation~\cite{Tolman1934,Oppenheimer:1939ne},
\begin{eqnarray}
{\di P(r)\over \di r}=-G_{\rm N}{[P(r)+\epsilon(r)]\left[M(r)+4\pi r^3P(r)\right]\over r^2-2G_{\rm N}\,rM(r)}.
\end{eqnarray}
Here, $G_{\rm N}=6.70\times10^{-45}\,{\rm MeV^{-2}}$ is the Newton's gravitational constant in natural unit and $M(r)$ is the core mass within radius $r$ that can be determined from
\begin{eqnarray}
{\di M(r)\over \di r}=4\pi r^2\epsilon(r).
\end{eqnarray}
With the known equation of state $\epsilon(r)=\epsilon(P(r))$, the differential equations are actually coupled equations of $P(r)$ and $M(r)$. There are two obvious initial conditions, $M(0)=0$ and $P(R)=0$, where $R$ is the radius of the neutron star. For practical calculations, it is more convenient to start with $M(0)=0$ and a given $P(0) (>0)$, work out $P(r)$, and find out the radius $R$ according to $P(R)=0$. 

Then, by following the discussions in Ref.~\cite{Regge:1957td,Hinderer:2007mb}, we are ready to calculate the tidal deformability $\Lambda$ of spherically symmetric neutron stars with the expression:
\bea
\Lambda={2\over3}k_2\left({R\over G_{\rm N}M}\right)^5.
\eea 
Here, the key quantity $k_2$ is the second tidal Love number whose expression was given by~\cite{Hinderer:2007mb}
\bea
k_2&=&{8C^5\over5}(1-2C^2)[2+2C(y-1)-y]\ \Big\{6C[2-y+C\times\nonumber\\
&&\!\!(5y-8)]+4C^3[13-11y+C(3y-2)+2C^2(1+y)]\nonumber\\
&&+3(1-2C^2)[2-y+2C(y-1)]ln(1-2C)\Big\}^{-1},
\eea 
where $C=G_{\rm N}M/R$ is the star’s compactness parameter and $y\equiv R H'(R)/H(R)$. To work with natural units, note that the Newton constant $G_{\rm N}$ is recovered in $C$ and in the following compared to those in Ref.~\cite{Regge:1957td,Hinderer:2007mb}.  Next, the unknown function $H(r)$ should be solved from the differential equation
\bea
&&H''+\left\{{2\over r}+G_{\rm N}e^\lambda\left[{2M(r)\over r^2}+4\pi r(P-\epsilon)\right]\right\}H'+\left[-{6e^\lambda\over r^2}\right.\nonumber\\
&&\left.+4\pi G_{\rm N}e^\lambda\left(5\epsilon+9P+{\epsilon+P\over C_{\rm v}^2}\right)-(\nu')^2\right]H=0,\label{Heq}
\eea
 where the derivatives are with respect to the radius $r$, the $r$-dependence of $P, \epsilon, \lambda,$ and $\nu$ is not shown explicitly, and $\lambda(r)$ and $\nu(r)$ are defined through the metric as $e^{\nu}=e^{-\lambda}=1-2G_{\rm N}*M(r)/r$~\cite{Regge:1957td}. The regularity of $H(r)$ around the center of neutron stars requires the boundary conditions $H(0)=H'(0)=0$ and $H''(0)\neq0$~\cite{Hinderer:2007mb}. Since for any constant $A$, the new function $A\,H(r)$ still satisfies Eq.\eqref{Heq} but would not alter the value of $y$, we can simply set $H''(0)=1$ for numerical calculations.

 \begin{figure}[!htb]
	\centering
	\includegraphics[width=0.42\textwidth]{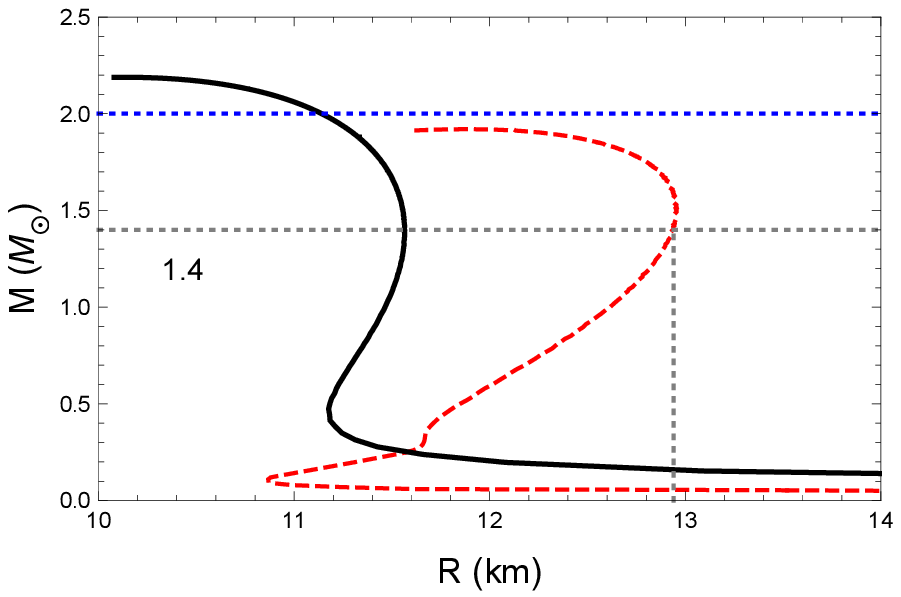}
	\includegraphics[width=0.42\textwidth]{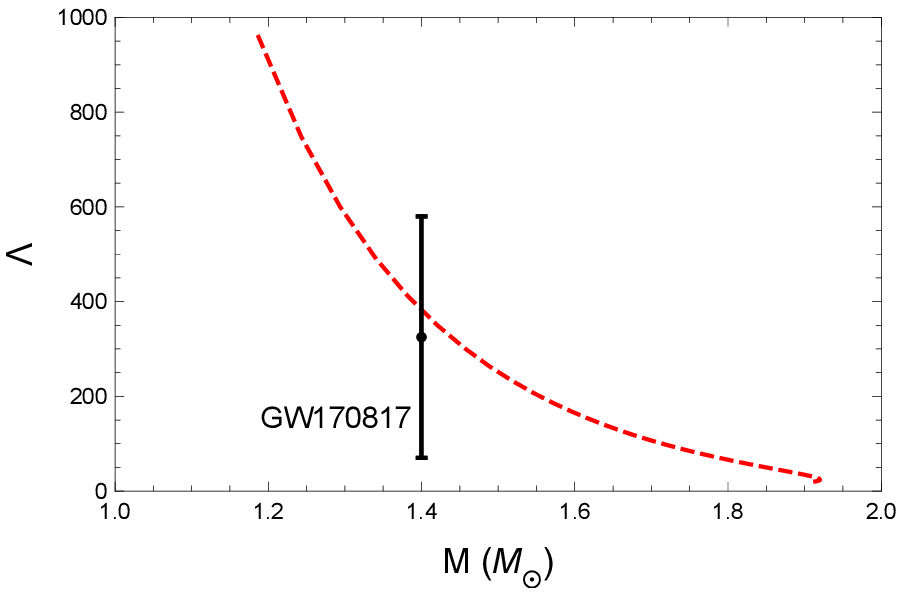}
	\caption{Upper panel: the mass-radius relations of quarkyonic neutron stars (red dashed line) and pure neutron stars (black solid line), lower panel: the tidal deformability of quarkyonic neutron stars (red dashed line) with the constraint from GW170817 (black error bar).}\label{MR}
\end{figure}
The numerical result for the mass-radius relation is illustrated together with that of pure neutron star~\cite{McLerran:2018hbz} in the upper panel of Fig.\ref{MR}. The inflection point around $0.3M_\odot$ corresponds to $n\sim 0.56\,n_0$ and indicates the emergence of quark degrees of freedom in the cores of quarkyonic neutron stars. Compared to that of pure neutron star, the EOS employed here is a bit too soft to generate the $2\,M_{\tiny \odot}$ mass of PSR J0740+6620~\cite{Fonseca2021} -- even when we vary $L_{\rm sys}(n_0)$ within the uncertainty. As softness is always expected for the quark system, the drawback implies that the dominance of quarks should be postponed to higher baryon density; and thus a peak structure is inevitable for the sound velocity and should be located at a super-saturation density for the QNM. Nevertheless, the radius of $1.4\,M_{\tiny \odot}$ neutron star is $12.94\,{\rm km}$, well within the updated constraints from {\rm NICER}, $R_{1.4}\in (11.39,13.10)\,{\rm km}$~\cite{Riley2021,Miller2021,Raaijmakers2021}. In the end, we present our prediction for the tidal deformability of quarkyonic neutron stars in the lower panel of Fig.\ref{MR} together with the constraint extracted from the multi-messenger event GW170817, $70\leq\Lambda(1.4M_{\odot})\leq580$~\cite{LIGOScientific:2018cki}. We are glad that the model well passes the test.

\section{Conclusions}\label{conclusion}
In this work, we extend our previous study of quarkyonic nuclear matter to quarkyonic neutron matter by introducing interactions with vector-isovector $\rho$ mesons. The three new parameters are fixed by fitting to $\rho$ meson mass in vacuum and the experimental constraints on the symmetry energy and its slope at saturation density. The features of chiral and vector condensates are obtained with increasing baryon density: while $\langle\sigma\rangle$ decreases slowly according to chiral symmetry restoration, $|\langle\omega_0\rangle|$ and $|\langle\rho_0^3\rangle|$ increases quickly due to the density interactions. Thermodynamic properties of the quarkyonic neutron matter are also explored, such as the pressure and sound velocity, the former of which is well consistent with experimental restrictions. Finally, the obtained equation of state is applied to solve the TOV equation for the quarkyonic neutron stars. It is a pity that the EOS is not stiff enough to give rise to the maximal two solar masses observed. The defect origins from the early dominance of quark degrees of freedom in our model (see the upper panel in Fig.\ref{dens_cond}) and is closely related to the absence of any peak in the sound velocity (see the lower panel in Fig.\ref{Pressure}). Nevertheless, the prediction of tidal deformability is well within the constraint extracted from the event GW170817 for the neutron star with mass $1.4M_{\odot}$.

In the future, the model should be improved by effectively suppressing quark fraction at baryon density $n_{\rm B}\lesssim5\,n_0$~\cite{Baym:2017whm}. Since the mean distance between two neutrons is still comparable to their own size at $n_{\rm B}=5\,n_0$, quarks' contribution must be secondary due to the preservation of strong confinement effect. At this point, it might be important to take into account gluon degrees of freedom which give rise to confinement, probably through a potential of Polyakov loop~\cite{Fukushima:2017csk}. However, such a kind of potential has never been reliably obtained in the first-principle lattice QCD simulations mainly due to the notorious sign problem. We hope the development of functional renormalization group and tensor network could help us to get an insight into the high density region.

\emph{Acknowledgments}---
G.C. is supported by the National Natural Science Foundation of China with Grant No. 11805290.

\end{document}